\documentclass[10pt,aps,prb,twocolumn]{revtex4}   

\usepackage{amsmath}    
\usepackage{amsfonts}  
\usepackage{amssymb}
\usepackage{graphicx}   
\usepackage{epstopdf}

\newcommand{\f}{\frac}

\newcommand{\be}{\begin{equation}}
\newcommand{\ee}{\end{equation}}
\begin{document}

\title{A blackbody is not a blackbox}
\author{Matteo Smerlak}\email{smerlak@cpt.univ-mrs.fr}
  \affiliation{Centre de Physique Th\'eorique\\ Campus de Luminy, case 907\\ 13288 Marseille cedex 9, France}

\date{\today}

\begin{abstract}\noindent
We discuss carefully the \emph{blackbody approximation}, stressing what it is (a limit case of radiative transfer), and what it is not (the assumption that the body is perfectly absorbing, i.e. \emph{black}). Furthermore, we derive the Planck spectrum without enclosing the field in a box, as is done in most textbooks. Athough convenient, this trick conceals the nature of the idealization expressed in the concept of a blackbody: first, the most obvious examples of approximate blackbodies, stars, are definitely not enclosed in boxes; second, the Planck spectrum is continuous, while the stationary modes of radiation in a box are discrete. Our derivation, although technically less elementary, is conceptually more consistent, and brings the opportunity to introduce to students the important concept of \emph{local density of states}, via the resolvent formalism. 

\end{abstract}

\maketitle

\begin{verse}
\begin{flushright}
\emph{``No colors anymore, I want them to turn black\\
$[...]$ I wanna see the sun blotted out from the sky\\
I wanna see it painted, painted, painted, painted black\\
Yeah!"\\} The Rolling Stones \end{flushright}
\end{verse}

\section{Introduction}

Planck's law for blackbody radiation is celebrated as a landmark in the history of physics, being the first physical law conflicting patently with classical mechanics. Derived in 1900, it is usually regarded as the igniter of the century of quantum mechanics. But its relevance goes beyond quantum mechanics.  Blackbody radiation is a striking example of universality in statistical mechanics -- the light radiated by a blackbody does not depend on its constitution, only on its temperature --, a topic which came to the foreground with the later studies of critical phenomena. More recently, Planck's distribution has appeared to fit with an astounding accuracy the cosmic microwave background (deviations are at most 50 per million\cite{CMB}), thus opening the era of precision cosmology.

Because of its seminal importance, one feels that the derivation of Planck's law presented to students should be as lucid as possible. In most textbook presentations\cite{landau,schrodinger,morse,miller,hecht,adkins,reif}, however, one apparently innocuous step undermines the understanding of its applicability: the radiation field is assumed to be enclosed in a box. This prompts students to think that the Planck spectrum is the one radiated by a \emph{black box}. This is true, of course, but also very misleading. If the blackbody radiation is really the blackbox radiation, why should stars, which are not enclosed in boxes, have a Planckian spectrum? Why should hot metal bars, bulb filaments, incandescent lava or the background of the universe have a Planckian spectrum?

Since the physics of thermal radiation generally does \emph{not} involve a box, we feel that the discussion of Planck's law should not either. In this note, we propose such a discussion. We introduce the blackbody approximation in the framework of radiative transfer theory, and derive the Planck spectrum using the notion of \emph{local} density of states. We do not claim originality in the method used, the resolvent formalism, which is standard in condensed matter physics and scattering theory. Rather, our aim is to promote an approach we believe to be pedagogically more transparent. (See also Ref. \cite{manybody} for a more advanced discussion of Planck's law from the many body theory perspective.)

The paper is organized as follows. In the next section, we discuss the blackbody approximation, and its relation with Kirchhoff's law. In section III, we define the local density of states, and derive Planck's law without the box, using the resolvent formalism. Section IV presents our conclusion.

\section{What is a blackbody?}
\subsection{The common definition}
A blackbody is usually defined as ``a body which completely absorbs all radiation incident on it". Landau and Lifschitz\cite{landau} add

\begin{quote}
Such a body can be realised in the form of a cavity with highly absorbing 
internal walls and a small aperture. Any ray entering through the aperture can 
return to it and leave the cavity only after repeated reflection from the walls 
of the cavity. When the aperture is sufficiently small, therefore, the cavity will 
absorb practically all the radiation incident on the aperture, and so the surface 
of the aperture will be a black body.
\end{quote}
Clearly, this picture makes perfect experimental sense. Theoretically, on the other hand, the nature of the idealization it is meant to express remains clouded. If a blackbody emits light, why is it called ``black"? And if it is an ideal emitter, why is it defined as an ideal ``absorber"? What is the role of the cavity? More importantly, what would an \emph{imperfect} blackbody be? A cavity which does not absorb all incident radiation? One with a bigger aperture?

\subsection{A continuous spectrum}

Many objects emit light like imperfect blackbodies. The most prominent among them is, of course, our Sun (see Fig. 1). Its spectrum, first measured in the early nineteenth century by Wollaston and Frauhofer, is well approximated by that of a blackbody of temperature between $5500$ and $6000$ K. It does displays deviations from it, the so-called Fraunhofer lines (corresponding to the absorption of certain frequencies by the solar atmosphere), but it is indisputable that, athough very dissimilar from an absorbing cavity with a tiny hole, the Sun qualifies as an imperfect blackbody. On the other hand, the spectroscopy of monoatomic gases shows clearly that their emission spectrum is \emph{not} Planckian. They emit only certain discrete frequencies (Fig. 2), determined by the electronic structure of the atoms.\footnote{Of course, a gas does have a continuous energy spectrum, corresponding to processes in which electrons are removed from their bound states with the nuclei -- photo-ionization. These, however, require higher energies, and this is why they are not easily seen with monoatomic gases. See Refs. \cite{Vollmer1,Nauenberg,Vollmer2} for a discussion of the transition from discrete to continuous spectra in gases.}

This observation hints at what a blackbody really is: \emph{a body with a rich energy spectrum, capable of exciting all frequencies of light by thermalization}.\footnote{This condition is salient in the case of the electromagnetic field because of its (quasi-)linearity: photons of different frequencies interact very little, and therefore cannot thermalize among themselves efficiently.}

From this perspective, the box definition appears paradoxical. As is well-known, a closed box selects certains light frequencies, through the condition $\omega_i=n\pi c/L_i$, where $L_i$ is the dimension of the box in the $i$-direction. Thus, instead of permitting a wide range of thermally excited frequencies, the box restricts the emission spectrum, even making it discrete. Of course, one could argue that the volume of the box can be made arbitrarily large, and therefore that this quantization of frequencies is not physically relevant. But this is precisely our point: as far as the frequencies of light are concerned, the box is not physically relevant.

\begin{figure}[h]
\begin{center}
\includegraphics[scale=0.08]{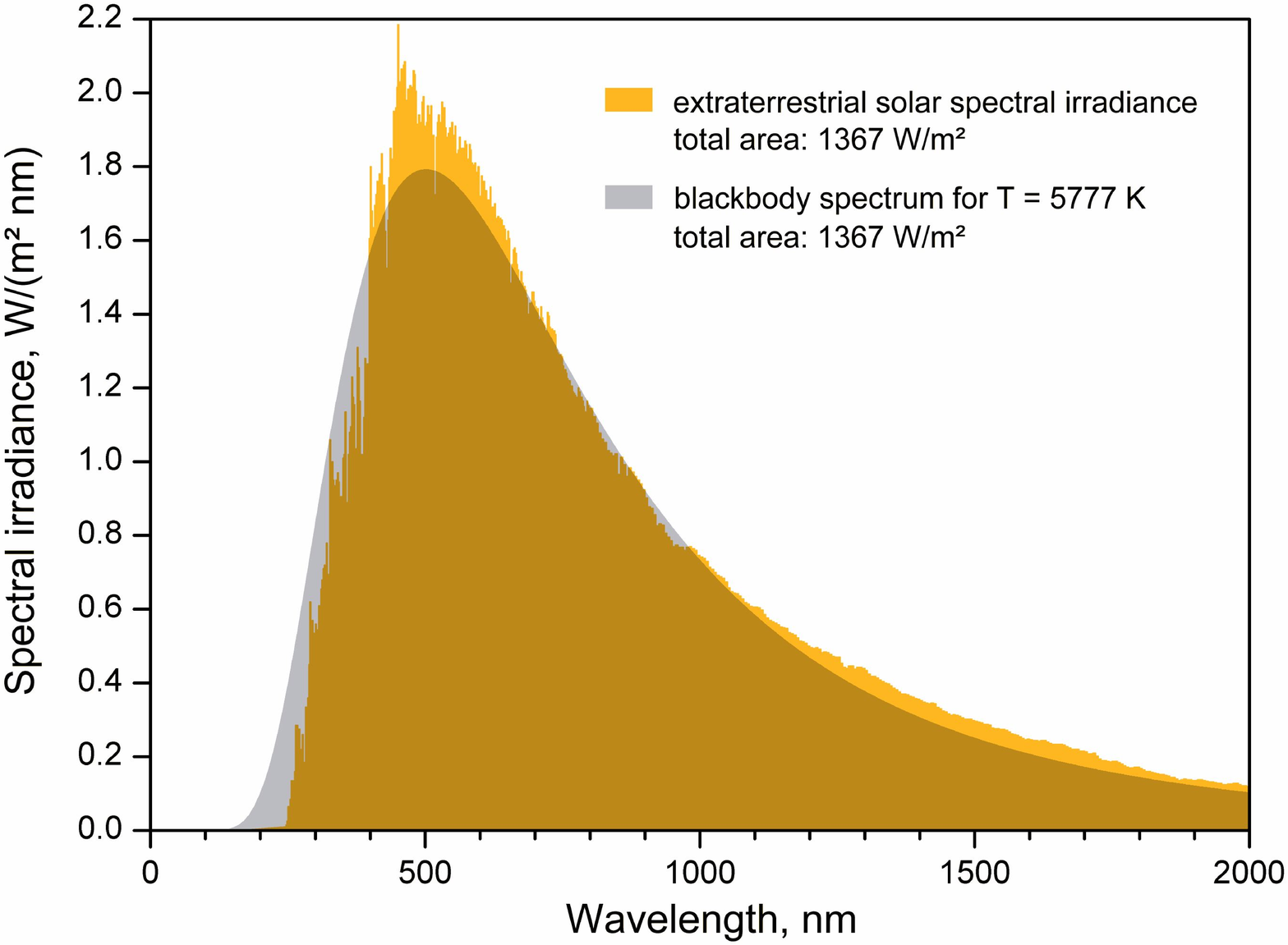}
\includegraphics[scale=0.25]{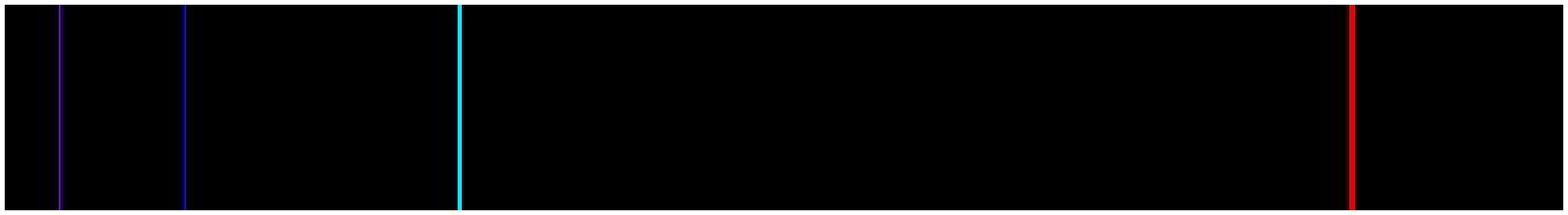}
\includegraphics[scale=0.25]{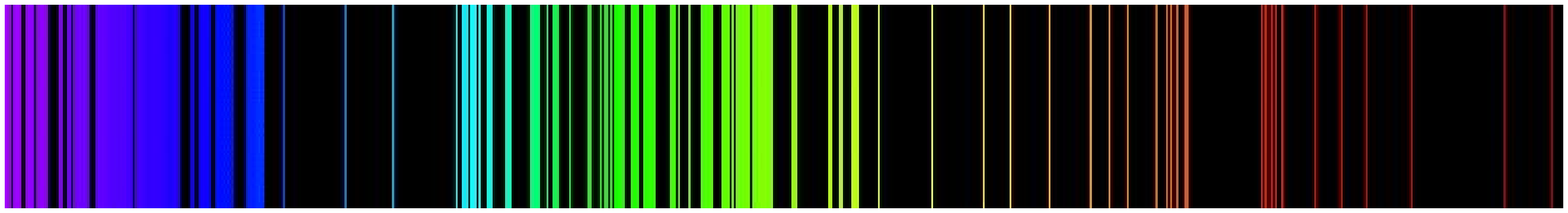}
\end{center}
\caption{The Sun is an imperfect blackbody, monoatomic gases are not (top: hydrogen; bottom: iron).}
\end{figure}

\subsection{Kirchhoff's law}

In its standard definition, a blackbody is one that ``absorbs all incident light'' -- a \emph{black} body. This fact alone should disturb the mindful student: how can a body be \emph{black}, and yet emit a \emph{colourful} spectrum of thermal light? In any case, what should the {\it absorptive} power of a body have anything to do with the caracteristics of its thermal {\it emission}?

The answer lies in an experimental observation which played a key role in the nineteenth century developments which led to Planck's successful analysis of thermal radiation, and which is too seldom mentioned in undergraduate discussions of thermal radiation -- Kirchoff's law.\cite{schirr} 

The radiative properties of a body are characterized by its \emph{emissivity} and \emph{absorptivity} (and scattering, which can usually be neglected). These can be defined by the following schematic model for the propagation of radiation within a medium.\cite{astro} As the (monochromatic) beam travels trough the medium, the variation of its energy density $u(l,\omega)$ receives two contributions: a positive one, corresponding to emission, and a negative one, corresponding to absorption:
\be\label{evol}
\f{du}{dl}(l,\omega)=\varepsilon(\omega)-\alpha(\omega) u(l,\omega).
\ee
The coefficients $\varepsilon(\omega)$ and $\alpha(\omega)$ are the emissivity and absorptivity of the body. Now, Kirchoff's law states that, although $\varepsilon(\omega)$ and $\alpha(\omega)$ largely depend on the constitution of the material, \emph{at thermal equilibrium, their ratio $J_T(\omega)\equiv\varepsilon(\omega)/\alpha(\omega)$ is universal}; it depends on temperature and frequency only. A good absorber ($\alpha(\omega)$ large) at a certain frequency is also a good emitter ($\varepsilon(\omega)$ large) at that frequency, and vice versa. 

At this point, our mindful student's worries should already be eased: if Kirchhoff's law is right, then blackbodies, which are by definition excellent absorbers, must also be excellent emitters. But further reflection should reveal a caveat in this line of thought: emissivity and absorptivity usually depend on the actual material used, while the blackbody radiation it emits does not. Why is that?

Some insight into this question is provided by the following consideration, due to Einstein.\cite{einstein} The interaction between matter and radiation boils down to transitions between energy levels: given two levels $a$ and $b$, with respective energies $E_a<E_b$, emission corresponds to an upgoing transition $a\rightarrow b$, while absorption correspond to a downgoing transition $b\rightarrow a$. The rate of these transitions, $\Gamma_{a\rightarrow b}$ and $\Gamma_{b\rightarrow a}$, is what controls at the microscopic level the absorptivity and emissivity of the body. Now, the condition of thermal equilibrium fixes the probabilites of each levels, through the Gibbs distribution $p_{a,b}\propto e^{-\f{E_{a,b}}{k_BT}}$. But, and this is the key point, it has no bearing on the transition rates $\Gamma_{a\rightarrow b}$ and $\Gamma_{b\rightarrow a}$ themselves, but {\it only on their ratio}. Indeed, in this microscopic perspective, thermal equilibrium translates into the condition of \emph{detailed balance}, according to which the probability flux between microstates cancel exactly:

\begin{equation}
p_a\Gamma_{a\rightarrow b}=p_b\Gamma_{b\rightarrow a}.
\end{equation}
This leads to 
\begin{equation}
\frac{\Gamma_{a\rightarrow b}}{\Gamma_{b\rightarrow a}}\propto e^{-\f{\hbar\omega_{ab}}{k_BT}},
\end{equation}
in which the RHS is a function of the transition frequency $\omega_{ab}$ and temperature only. Kirchhoff's law is a consequence of this constraint on the transition rates imposed by detailed balance.\footnote{In fact, Einstein \emph{derived} the Bose-Einstein distribution from detailed balance, understanding that both spontaneous and stimulated emission are required for the coexistence of thermal radiation and thermal matter.\cite{lewis}}




Combining the phenomenological radiative transfer equation (\ref{evol}) with Einstein's microscopic model, we understand that thermal equilibrium of the material source, through the condition of detailed balance, constrain the ratio of emission to absorption -- Kirchhoff's law -- but not their respective rates independently: this is why different materials have different thermal emission and absorption properties. It is only under a {\it further assumption} that a universal function describing blackbody emission -- Planck's function -- can be obtained. 

What is this further assumption? Is it that the body ``absorb all light incident on it", as in the standard definition? In other words, that the body be perfectly absorbing (`black') on the whole spectrum? No! Such a material does not exist.\footnote{For the emissivity to not depend on the frequency, Thomson scattering should be the dominant process in the matter-radiation interaction, and this {\it cannot} be the case near thermal equilibrium.} That is, the condition of `blackness' entering the standard definition of a blackbody {\it cannot} be the idealization underlying blackbody radiation. Then what is it?

\subsection{Optical thickness}

The answer should be obvious by now: the additional condition is that the radiation field {\it itself} should be at thermal equilibrium. This means that the random processes of emission and absorption of light by the hot body should have reached their stationary state. Solving (\ref{evol}) with the initial condition $u(l=0,\omega)=0$, we find

\be
u(l,\omega)=(1-e^{-\alpha(\omega) l})J_T(\omega)
\ee
In particular, we see that if $\alpha(\omega) l\gg 1$,  the so-called `optically thick' limit, the observed energy density is given by $J_T(\omega)$, which is nothing but the Planck spectrum. In other words, the coefficient $\alpha(\omega)$ measures the rate of convergence of the emitted light to its equilibrium value $J_T(\omega)$ -- the blackbody approximation is just the condition that $\alpha(\omega) l\gg 1$. This is achieved not only for good absorbers ($\alpha(\omega)$ large), but also for objects involving long optical paths ($l$ large). 

This is precisely what a closed cavity does: it provides the conditions for this convergence process to take place, whatever the intrinsic properties $\varepsilon$ and $\alpha$ of the material within the cavity. The box, in other words, is a useful expedient for the actual production of thermal radiation in the laboratory {\it when the hot body is not black}: it permits to reach the optically thick limit even if the absorptivity is low, because it generates very long paths $l$ withing the material.\footnote{This explains Landau and Lifschitz's otherwise cryptic observation that the blackbody is not the cavity itself, but the ``surface of [its] aperture".} But the presence of a box is of course not a necessary condition for the thermalization of light: stars too are optically thick -- and this is why their spectrum looks Planckian.



\section{Planck's law without the box}

\subsection{Partition of energy}

From the previous discussion, we know that the {\it equilibrium} spectral energy density coincides with the function $J_T$ in Kirchoff's law: the energy contained in thermal radiation within an infinitesimal volume $d^3x$ about a point $x$ in the frequency range $d\omega$ is
\be
dE(x,\omega)=J_T(x,\omega)d^3xd\omega
\ee

How can one compute this function? The answers to this question is obvious when one recalls that the total energy of the field is distributed over its modes, and that modes with the same frequency $\omega$ are degenerate, i.e. store the same energy. Hence:
\begin{align}
dE(x,\omega)= & \ (\textrm{energy of each $\omega$-mode})\nonumber\\&\times(\textrm{occupation number of each $\omega$-mode})\nonumber\\&\times(\textrm{number of $\omega$-modes accessible at $x$})
\end{align}

From quantum mechanics, we know that the energy stored in a mode with frequency $\omega$ is $E(\omega)=\hbar\omega$. Moreover, the occupation number $n_T(\omega)$ of each such mode at thermal equilibrium is given by the Bose-Einstein distribution, 

\be
n_T(\omega)=\f{1}{e^{\f{\hbar\omega}{k_BT}}-1}.
\ee
It follows that the thermal energy density $J_T(x,\omega)$ is given by 

\be\label{planck}
J_T(x,\omega)=\hbar\omega\left(\f{1}{e^{\f{\hbar\omega}{k_BT}}-1}\right)\rho(x,\omega),
\ee
where $\rho(x,\omega)$ denotes the density of modes with frequency $\omega$ around the point $x$. Indeed, for an observer localized at $x$, certain modes might be inaccessible, or only partly accessible: $\rho(x,\omega)$ is the \emph{space-resolved} density of modes, often called (using quantum mechanical parlance) the \emph{local density of states} (LDOS). Of course, if the medium is homogeneous, this quantity does not depend on $x$. However, in more general situations (such those described at the end of the next section), it does.

\begin{figure}[h]
\centering
\includegraphics[scale=0.3]{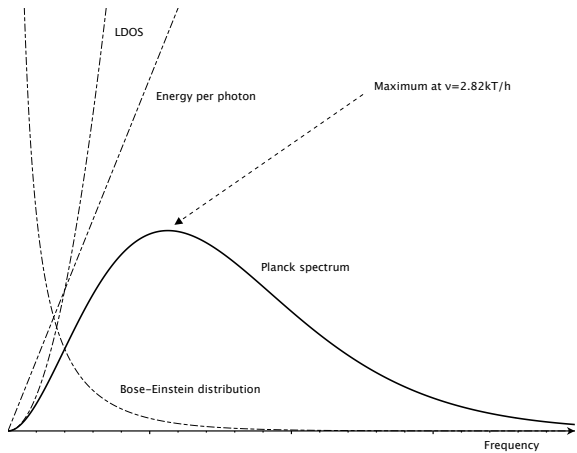}
\caption{Balance between the LDOS, the energy per mode and the Bose-Einstein distribution.}
\label{balance}
\end{figure}

All in all, the Planck spectrum can be described as the outcome of the balance between the energy per mode and the LDOS, which tend to grow as $\omega$ gets large, and the Bose-Einstein distribution, which favors the low-frequency modes. This is illustrated in Fig. \ref{balance}.
\bigskip

\subsection{A hint of spectral theory}

The reason why the evaluation of the LDOS is non-trivial mathematically is because the light frequencies span a \emph{continuous} spectrum. Had it been discrete, the LDOS would just have been the degeneracy of $\omega$, understood as an eigenvalue of the wave operator. In the continuous case, however, the notion of degeneracy is subtler, for it involves the definition of a \emph{measure} on the spectrum. This is analogous to the ambiguity one faces when trying to extend a discrete sum to an integral: the appropriate weighting of the integration variable is not obvious anymore. The case of the standard measure on the sphere, which involves a factor $\sin\theta$, is an obvious example of such subtleties. Of course, in this case, the appropriate measure is determined by the condition of rotational invariance. What determines the spectral measure of an operator?

A powerful tool to answer such a question is the `resolvent formalism'.\cite{teschl} Given a (self-adjoint) operator $A$, the resolvent is the operator-valued function of the complex variable $z$ defined by

\be
R(z)=(A-z)^{-1}.
\ee

By construction, $R(z)$ is analytic in $\mathbb{C}\setminus\sigma(A)$, the complement of the spectrum of $A$. The discrete part of $\sigma(A)$ corresponds to isolated poles of $R(z)$, while the continuous part of $\sigma(A)$ generates a branch cut along the real axis. In other words, the spectrum of $A$ is encoded in the analytic structure of $R(z)$.

Given any two vectors $\psi$ and $\phi$, it follows from the spectral theorem that

\begin{align}\label{spectral}
\langle\varphi\vert R(z)\psi\rangle&=\int_{\sigma(A)}\f{d\mu_{\varphi,\psi}(\lambda)}{\lambda-z}\\\nonumber&=\sum_{\lambda\in\sigma_d(A)}\f{g(\lambda)}{\lambda-z}+\int_{\sigma_c(A)}\f{\rho_{\varphi,\psi}(\lambda)d\lambda}{\lambda-z}
\end{align}
where $d\mu_{\varphi,\psi}$ is the spectral measure associated to $\varphi$ and $\psi$, and in the second equality the spectrum is decomposed into its discrete $\sigma_d(A)$ and continuous $\sigma_c(A)$ parts.\footnote{We disregard subtleties related to singular spectra.} The spectral density $\rho_{\varphi,\psi}(\lambda)$ is therefore defined as the density of the continuous part of the spectral measure $d\mu_{\varphi,\psi}$ with respect to the Lebesgue measure $d\lambda$.

When $A$ is a wave operator, and $(\varphi,\psi)=(x,y)$ are position (generalized) eigenvectors, the quantity $G(x,y;z)=\langle y\vert R(z) x\rangle$ is called in the physics literature the `Green function'. The spectral decomposition of its diagonal elements $G(x,x;z)$ reads

\be\label{diagonal}
G(x,x;z)=\sum_{\lambda\in\sigma_d(A)}\f{g(\lambda)}{\lambda-z}+\int_{\sigma_c(A)}\f{\rho_{x}(\lambda)d\lambda}{\lambda-z}.
\ee
This expression provides the spectral density $\rho_x(\omega)$ with the following interpretation: the generalized eigenvalue $\lambda$, when analyzed through a state localized at the point $x$, comes with a weight $\rho_x(\lambda)$. This weight is the overlap of the density of $\lambda$-eigenmodes with the local state $\vert x\rangle$ -- in other words, the LDOS $\rho(x,\lambda)$ of $A$. 

Moreover, the formula (\ref{diagonal}) indicates a procedure to evaluate $\rho_x(\lambda)$. Indeed, just as the degeneracy $g(\lambda)$ of an isolated eigenvalue $\lambda$ can be read off as the residue of $\langle y\vert R(z) x\rangle$ at $z=\lambda$, the LDOS is given by the discontinuity of $G(x,x;z)$ along the branch-cut singularity. This is the so-called Stieltjes-Perron inversion formula (Appendix A):
\be\label{LDOS}
\rho(x,\lambda)=\f{1}{2i\pi}\lim_{\epsilon\rightarrow 0}\Delta_{\epsilon} G(x,x;\lambda),
\ee
with 
\be
\Delta_{\epsilon} G(x,x;\lambda)=G(x,x;\lambda+i\epsilon)-G(x,x;\lambda-i\epsilon).
\ee

\subsection{Resolvent of the Laplace operator}

Let us apply this resolvent formalism to the case of electromagnetism. For the sake of simplicity, we shall consider here the scalar Helmholtz equation -- the two independent polarizations of the field will be taken into account \emph{a posteriori}, by multiplying the LDOS thus obtained by a factor of $2$. 

The Helmholtz equation in vacuum for a monochromatic wave $\phi_{\omega}$ reads
\be
-\Delta\phi_{\omega}(x)=\f{\omega^2}{c^2}\phi_{\omega}(x),
\ee
which is an eigenvalue equation for the Laplacian, with eigenvalue $\lambda=\omega^2/c^2$. We introduce the resolvent $(-\Delta-z)^{-1}$, and evaluate the corresponding Green function $G(x,y;z)$ by Fourier transform

\be
G(x,y;z)=\int \f{d^3k}{(2\pi)^3}\f{e^{ik.(x-y)}}{k^2-z}.
\ee

\subsection{The vacuum LDOS}

The two integrals $G(x,y;\lambda\pm i\epsilon)$ are readily evaluated by residue calculus (Appendix B), yielding
\be\label{disc}
\lim_{\epsilon\rightarrow 0}\Delta_{\epsilon} G(x,x;\lambda)=\f{i\sqrt{\lambda}}{2\pi}=\f{i\omega}{2\pi c}.
\ee

Using (\ref{LDOS}) and the fact that $d\lambda=2\omega/c^2d\omega$, we obtain the electromagnetic vacuum LDOS (with a factor of $2$ for the two polarizations):

\be
\rho(x,\omega)=\f{\omega^2}{\pi^2c^3}.
\ee

Note that, as expected, the translational invariance of the vacuum translates into the independence of the LDOS on the space point $x$. Plugging this value into (\ref{planck}) immediately yields Planck's law,

\be
J_T(\omega)=\f{\hbar\omega^3}{\pi^2c^3}\f{1}{e^{\f{\hbar\omega}{k_BT}}-1}.
\ee

\subsection{Surface effects}

\begin{figure}[h]
\centering
\includegraphics[scale=0.3]{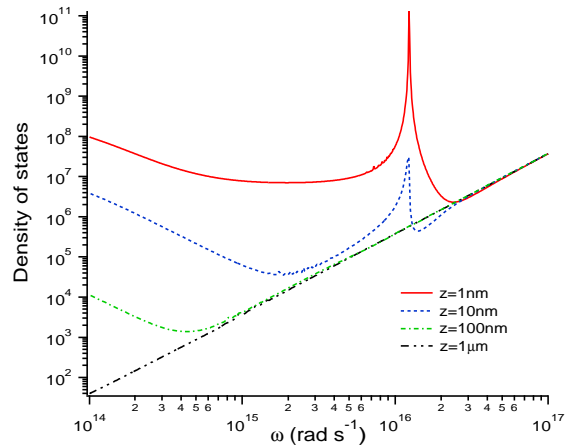}
\caption{LDOS versus frequency at different heights above a semi-infinite sample of aluminium, from Joulain \emph{et al.}\cite{ldosinterface} The resonance around $10^{16}$ rad/s corresponds to the excitation of surface-plasmon polaritons.}
\end{figure}

We conclude this section by an illustration of the strength of the LDOS method for a finer analysis of thermal radiation, in particular close to a metallic surface. At a distance to the hot body comparable to the thermal wavelength, the vacuum approximation breaks down, as the field reveals evanescent modes and polariton excitations. Such local excitations have been shown to modify significantly the LDOS, and hence the hot body spectrum, notably by enhancing monochromaticity (Fig. 3), spatial coherence, and directivity.\cite{ldosprl,ldosinterface,ldosnature} This shows another facet of the blackbody approximation which is not usually emphasized: it is a \emph{far-field} approximation.

\section{Conclusion}

In this paper, we have tried to disentangle two aspects of the standard picture of a blackbody as a cavity with a small aperture. The first one relates to the efficiency of the thermalization of light through its almost everlasting interaction with the walls of the cavity: this is the one invoked in Landau and Lifschitz's definition. Although it does not make the conditions of thermalization of radiation explicit, the cavity picture is useful to demonstrate how a blackbody can be realized in the laboratory. The second aspect relates to the selection of certain frequencies, and the subsequent Fourier space mode-counting argument. Unlike the former, this aspect does not merely serve illustrative purposes, but usually enters the actual derivation of the Planck spectrum. As such, it appears to students as an important feature of thermal radiation. We have argued that it is not, and that it actually contradicts a key feature of blackbody radiation, namely the fact that \emph{all} wavelengths are emitted.

Our approach, focused on local properties of the electromagnetic field, allows to derive Planck's law without using a blackbox, and provides tools for subtler considerations, such as the near-field regime of thermal radiation. With this perspective, we believe that students are less likely to miss the point of the blackbody approximation, which expresses an idealization of the \emph{interactions} between the electromagnetic field and heated materials -- and not of the materials themselves.

\begin{acknowledgments}
I am grateful to the African Institute for the Mathematical Sciences (where this work was started) for giving me, and many others, the opportunity to do science in Africa. I also thank Mohammed Suleiman Hussein Suleiman for discussions on Planck's law and much more, and all the students of the 2007 AIMS promotion for their enthusiasm and generosity.
\end{acknowledgments}

\appendix
\section{The Stieltjes-Perron formula}
Let a complex function $f$ be defined as 
\be
f(z)=\int_Idt\ \f{\rho(t)}{t-z}
\ee
where $\rho$ is a continuous density on the open interval $I$. We consider the values $f^{\pm}\equiv f(\lambda\pm i\epsilon)$, for $\lambda\in I$, and use the well-known identity (the Sokhatsky-Weierstrass theorem)
\be\label{sktheorem}
\lim_{\epsilon\rightarrow 0}\ \f{1}{t-\lambda\mp i\epsilon}=\mathcal{P}(\f{1}{t-\lambda})\pm i\pi\delta(t-\lambda)
\ee
where $\mathcal{P}(\f{1}{t-\lambda})$ is the Cauchy principal value. It follows that
\begin{eqnarray}
\lim_{\epsilon\rightarrow 0}\ (f^{+}-f^{-})& = &\int_I dt\ \rho(t)2i\pi\delta(t-\lambda)\\ & = &2i\pi\rho(\lambda),
\end{eqnarray}
which is the Stieltjes-Perron inversion formula.

\section{Evaluation of $G(x,y;\lambda\pm i\epsilon)$}

Consider the two integrals
\be
G(x,y;\lambda\pm i\epsilon)=\int \f{d^3k}{(2\pi)^3}\f{e^{ik.(x-y)}}{k^2-\lambda\mp i\epsilon},
\ee
denoted $G^{\pm}$ for simplicity. The angular integration is straightforward, and yields
\be
G^{\pm}=\f{2\pi}{(2\pi)^3}\f{2}{\vert x-y\vert}\int_0^{\infty}dk\ \f{k\sin(k\vert x-y\vert)}{k^2-\lambda\mp i\epsilon}.
\ee
The remaining integral is evaluated using the residue theorem, considering as integration contours the standard half-circles $\gamma_{\pm}$, enclosing the poles $k_{\pm}=\sqrt{\lambda\pm i \epsilon}$ respectively. We obtain
\be
G^{\pm}=\pm\f{i}{4\pi}\f{\sin(k_{\pm}\vert x-y\vert)}{\vert x-y\vert},
\ee
The discontinuity across the cut is then given by
\be
\lim_{\epsilon\rightarrow 0}(G^+-G^-)=\f{i}{2\pi}\f{\sin(\sqrt{\lambda}\vert x-y\vert)}{\vert x-y\vert},
\ee
from which (\ref{disc}) follows by setting $y=x$.


\end{document}